\documentstyle[multicol,aps,prl,epsfig]{revtex}

\begin{document}
\title{Dynamic Scaling in Diluted Systems Phase Transitions: Deactivation trough
Thermal Dilution}
\author{Manuel I. Marqu\'es and Julio A. Gonzalo}
\address{Dpto. de F\'{i}sica de Materiales C-IV\\
Universidad At\'onoma de Madrid\\
Cantoblanco 28049 Madrid Spain\\
email: julio.gonzalo@uam.es}
\date{29-8-2000}
\maketitle

\begin{abstract}
Activated scaling is confirmed to hold in transverse field induced phase
transitions of randomly diluted Ising systems. Quantum Monte Carlo
calculations have been made not just at the percolation threshold but well
bellow and above it including the Griffiths-McCoy phase. A novel
deactivation phenomena in the Griffiths-McCoy phase is observed using a
thermal (in contrast to random) dilution of the system
\end{abstract}



\vskip1pc

\begin{multicols}{2}
\narrowtext
\parskip=0cm

The presence of quenched disorder in quantum phase transitions at zero
temperature is a topic of current interest. Two of the most important
special properties of the disordered quantum phase transitions are the
appearance of {\bf activated dynamic scaling} and the existence of {\bf %
Griffiths-McCoy singularities} \cite{griffiths,mccoy} even away from the
critical point.

Activated dynamic scaling was first analytically proved to hold in the
disordered one dimensional Ising model in a transverse field \cite{fisher}
and there are many other results for this model in the literature \cite
{shankar,mccoyII,mckenzie,igloi,young}. The disordered Ising model in a
transverse field considered as a quantum spin glass provides a reasonable
description of the system $LiHo_{x}Y_{1-x}F_{4}$ \cite{wu} and also,
considering the existence of long-range correlated disorder in it, may be an
appropriate model to describe non-Fermi liquid behavior in certain $f$%
-electron systems \cite{castro,andrade,rieger}. Activated dynamic scaling
implies the existence of an infinite dynamic critical exponent $(z=\infty $%
). It means that instead of the typical power law relationship between the
characteristic time scale and the characteristic length scale, $\xi _{\tau
}\sim \xi ^{z},$ a new exponential relation appears $\xi _{\tau }\sim
exp(const\times \xi ^{\psi })$ with $\psi =1/2$ for the one-dimensional
Ising model \cite{fisher}. There are nearly no analytical results for higher
dimensions. The activated dynamic scaling seems to disappear for
two-dimensional \cite{riegerII} and three dimensional Ising glass systems 
\cite{guo}, but it has been proved to hold in the disordered two-dimensional
Ising model in a transverse field by means of Quantum Monte Carlo
simulations \cite{pich,riegerIII} and renormalization group analysis \cite
{motrunich}. The only analytical prediction of activated scaling in
dimensions higher than one has been made for the dilution probability
transition at the percolation threshold $(p_{c})$ of a diluted Ising system
in a transverse field \cite{senthil}, where percolation critical exponents
have been found.

The phase boundary of this model at $T=0$ was studied a long time ago \cite
{harris,stinchcombe,bhattacharya} and is expected to have a multi-critical
point and a straight vertical phase boundary separating the ferromagnetic
phase from the quantum Griffiths phase at the percolation threshold. The
existence of this boundary and the activated scaling predicted by Senthil et
al. has been recently checked by means of Quantum Monte Carlo simulations,
comparing the dilution probability transitions at constant transverse fields
for $p$ values bellow an above the one corresponding to the
multicritical point\cite{ikegami}.

However, there is no study of the existence of activated scaling for
transitions tuned by the transverse field in diluted Ising systems. These
systems present a great advantage: fixing the probability of occupied places 
$(p)$ and varying de transverse field $(\Gamma )$ at $T=0,$ it is possible
to follow the evolution of the activated scaling at zero temperature in the
ferromagnetic phase $(p>p_{c})$ at the phase boundary $(p=p_{c})$ and in the
Griffiths phase $(p<p_{c})$. In this paper we will use the diluted Ising
model in a transverse field to directly determine the existence and
evolution of activated scaling in the three different zones of the phase
diagram. We will show how the activated scaling exists not just at the
percolation threshold but well above it and how it reaches a maximum near
the phase boundary $(p=p_{c}$) keeping mostly constant all the way through
the Griffiths zone. We start from the two-dimensional $(d=2)$ diluted Ising
model in a transverse field with Hamiltonian given by 
\begin{equation}
{\cal H}=-\sum\limits_{\left\langle i,j\right\rangle }\varepsilon
_{i}\varepsilon _{j}\sigma _{i}^{z}\sigma _{j}^{z}-\Gamma
\sum\limits_{i}\varepsilon _{i}\sigma _{i}^{x}  \label{isingtrans}
\end{equation}
where $\varepsilon _{i}\in \left\{ 0,1\right\} $ are quenched random
variables producing the dilution. $\varepsilon _{i}=1$ with probability $p$
and $\varepsilon _{i}=0$ with probability $1-p$. In order to perform the
Monte Carlo calculations we use the Suzuki-Trotter decomposition \cite
{suzuki} and we map the phase transition at $T=0$ of the two-dimensional
quantum system in a three dimensional classical system with action 
\begin{equation}
{\cal S}=-K_{hor}\sum\limits_{\tau ,\left\langle i,j\right\rangle
}\varepsilon _{i}\varepsilon _{j}s_{i}(\tau )s_{j}(\tau
)-K_{ver}\sum\limits_{\tau ,i}\varepsilon _{i}s_{i}(\tau )s_{i}(\tau +1)
\label{suzukidescom}
\end{equation}
with $K_{hor}=\Delta \tau $, $K_{ver}=-(1/2)ln\left[ tanh(\Delta \tau \Gamma
)\right] $ and $\Delta \tau \rightarrow 0$. This limit may be taken into
account exactly considering a continuous time algorithm \cite{riegerIII},
however making use of the universality between the model with $\Delta \tau
\rightarrow 0$ and the model with $\Delta \tau \neq 0$, we may simulate the
usual discrete Ising model, but with anisotropic interactions. In order to
avoid problems arising from de critical slowing down at the critical point
we use the Wolff single cluster algorithm \cite{wolff}. It is important
to notice that once an spin is diluted, the whole imaginary time axis
(Trotter axis) is also diluted, which means that the random disorder in a
quantum system is equivalent to the long range correlated disorder in
classical systems \cite{weinrib} giving rise to a different universality
class. In order to obtain the $T=0$ phase boundary we fix the occupied site
probability $(p)$ and we consider different values of the transverse field $%
(\Gamma )$ at intervals of $\Delta \Gamma =0.02$. We really do not fix the
probability but the concentration, it means that we do not use a grand
canonical distribution of the vacancies but a canonical distribution,
however both kinds of constrains are supposed to belong to the same
universality class \cite{aharony}.

The critical magnetic field for each probability $\Gamma (p)$ is determined
by a method previously used for Ising spin glasses \cite{riegerII,guo}.
First it is necessary to compute the Binder Cumulant average from a certain
number of realizations of the disorder (in our case we consider $500$
different realizations), this is done for a fixed value of ($\Gamma ,p)$ and
for a fixed value of the size (we consider the values $L=8,12,16,24$). Then
we study the evolution of such a cumulant for different sizes of the Trotter
axis (we consider up to $L_{\tau }=600$). Due to the dynamic scaling form of the
Binder Cumulant it has a peak as a function of $L_{\tau}$. At the critical
point ($\Gamma _{c}(p))$ the peak height is independent of $L$ and the
values of $L_{\tau }$ at the maximum, $(L_{\tau })_{m}$, vary as $L^{z}$ for
conventional dynamic scaling and like $(L_{\tau })_{m}\sim exp(const\times
L^{\psi })$ for activated scaling. Of course, all the non-universal
quantities, as for example the critical values of the transverse field, will
depend on the model and in particular on $\Delta \tau $. In the present work
we choose $\Delta \tau =1/5$.

The results for the phase boundary at $T=0$ are shown in Fig.1 together with
the phase boundary obtained for the classical system by means of
conventional scaling. The points where the existence of activated scaling is
going to be checked are $p=1$ (pure case), $p=0.8,0.7$ ($p>p_{c}$), $%
p=0.6\simeq p_{c}\simeq 0.59$ \cite{ziff} and $p=0.55,0.5$ ($p<p_{c}$)
(Griffiths phase). The results for $(L_{\tau })_{m}$ vs. $L$ are presented
in Fig.2. Note how the scaling is not activated for the pure case $(p=1)$
where the dynamical exponent is found to be $z\approx 1$, but it starts to
activate as the dilution is increased. To ensure that for diluted cases the
scaling is activated Fig.3 presents a plot of $ln(L_{\tau })_{m}$ vs. $L$.
The straight line behavior for the values $p\neq 1$ clearly indicates that the scaling
is activated. The evolution of the values $\psi (p)$ is presented in the
inset. Note how it grows monotonically until $p\simeq p_{c}$ and then it
keeps approximately constant inside the Griffiths zone.

\begin{figure}
\epsfysize=7cm\epsfxsize=\columnwidth\epsfbox{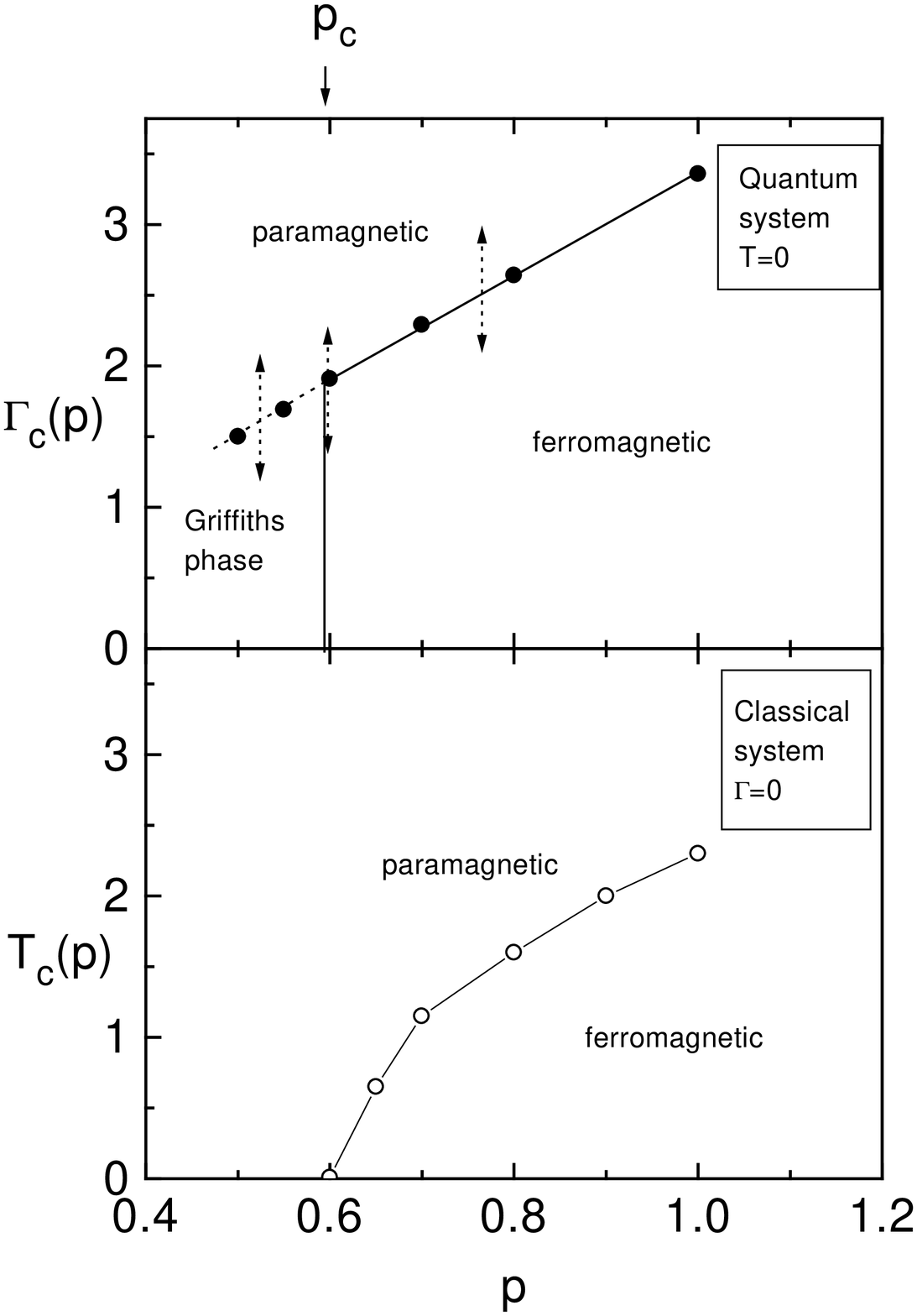}
\caption{Phase diagram calculated for the classical model, $T_{c}$ vs. $p$ (white points)
and for the quantum model at $T=0$, $\Gamma_{c}$ vs. $p$ (black points). Solid lines
are a guide for the eye. Dotted line indicates the phase boundary
in the Griffiths region ($p<p_{c}$). The three arrows indicate the transitions to study, above the
percolation thershold, at the percolation thershold, and inside the Griffiths zone.}
\label{fig1}
\end{figure}

\begin{figure}
\epsfysize=7cm\epsfxsize=\columnwidth\epsfbox{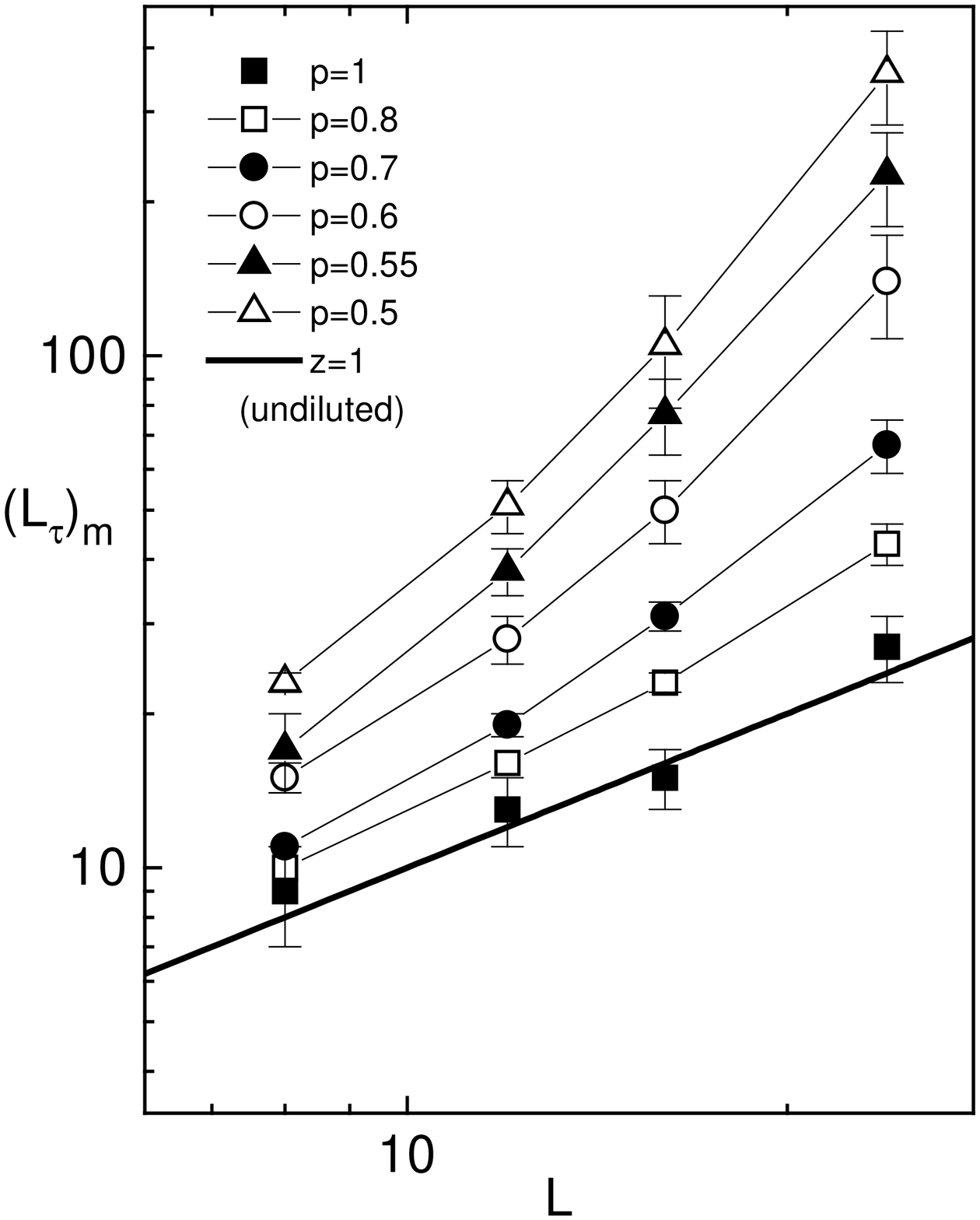}
\caption{$(L_{\tau })_{m}$ vs. $L$ for different values of the spin concentration.
Thin straight lines are just a guide for the eye. Thick line represents the behavior
expected for the pure system ($z=1$).}
\label{fig2}
\end{figure}

\begin{figure}
\epsfysize=9cm\epsfxsize=\columnwidth\epsfbox{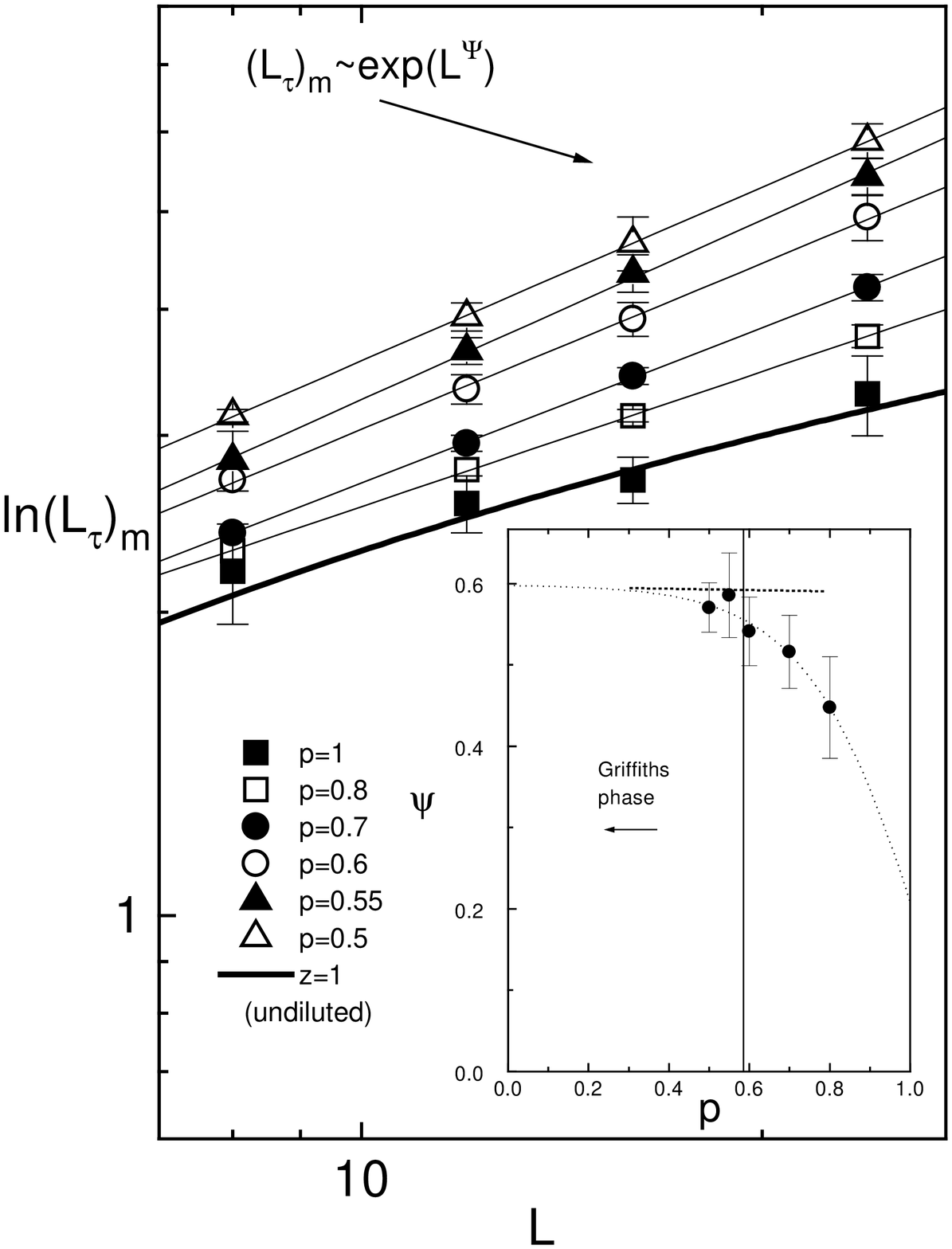}
\caption{$ln(L_{\tau })_{m}$ vs. $L$ for different values of the spin concentration.
Thin lines are the linear fittings and the thick line is the behavior expected for
the pure system ($z=1$). Inset shows the behavior of the exponent $\psi (p)$. Dotted line
is a sigmoidal fitting.}
\label{fig3}
\end{figure}

So, basically, the results presented indicate that the scaling is activated
not just near the percolation threshold but well above it, and that it keeps
nearly constant when the transition considered is inside of the
Griffiths-McCoy quantum zone, at least for the values of $p$ considered,
which are near $p_{c}$ (it may be also a maximum). Note that we are using, up to now, a {\bf purely
random} procedure to produce the diluted samples.

The existence of the Griffiths-McCoy zone is due to rare regions which are
locally in the wrong phase. Basically there are some strongly coupled
regions or clusters in the ferromagnetic phase even for spin concentrations
smaller than the percolation threshold of the system. This zones are found
studying the tail of local susceptibilities \cite
{igloi,youngII,igloiII,riegerIV,thill,riegerV,guoII}. The quantum
transitions allow the existence of these clusters in the wrong phase but
they are very difficult to detect in classical systems. As we have shown the
phase transitions (tuned by the transverse field) due to these rare regions
also present activated scaling.

However there might be a possible way to deactivate the scaling in the
Griffiths zone. Using the Suzuki-Trotter formalism the existence of the
wrong phase comes from clusters which are infinite in the imaginary time
direction, but they still are {\bf finite} in the perpendicular slices. That
is the reason why is possible to ''feel'' the effect of the dilution by the existence of an activated scaling. However, if the dilution
is not introduced randomly, but incorporating an {\bf infinite correlation
length} between spins and vacancies some of the clusters formed in the
slices may be considered to have {\bf infinite size}. The wrong phase
arising in these clusters is equivalent to the pure system ordered phase and
is not affected by disorder. In this case the system in the Griffiths zone
is expected to show deactivated scaling. Of course, this will happen just if
the phase transition is due to the existence of strongly coupled regions
(i.e. if the system is in the Griffiths phase $(p<p_{c})$), but if the
magnetic response is due to whole system (i.e. $p>p_{c}$) the introduction
of long range correlated disorder will affect changing the universality
class of the system (as in thermal classic transitions \cite{weinrib}) and
enhancing the activation of the scaling \cite{rieger}.

A way to produce this kind of dilution is using the {\bf thermal} dilution
instead of a random dilution \cite{marques}. With this kind of dilution the
system belongs to the universality class of {\bf long-range correlated
disordered systems} with an exponent $a=2-\eta $ \cite{marquesII}, being $%
\eta $ equal to $0.25$ for the classical ($d=2)$ Ising model (the way to
produce thermal dilution is described in \cite{marques}). Basically the
procedure is as follows, the pure system is first thermalized to criticality
and then one kind of spins are turned into vacancies. The samples produced
by thermal dilution at criticality will have an spin concentration near $%
p=0.5$, so we can compare them only with samples diluted randomly with
probability $p=0.5$. Both kinds of dilution will be at the Griffiths phase
since in both cases $p_{c}(a)>0.5$ \cite{prakash}.

The analysis performed has been exactly the same as before but now going up
to $L=40$. By the Binder Cumulant we have been able to determine the
critical transverse magnetic field ($\Gamma _{c}(thermal)\simeq 3.13$) and
the relation between $(L_{\tau })_{m}$ and $L$. Fig.4 compares de behavior
of the thermal dilution with the behavior of the random dilution with $p=0.5$%
. The thick line represents the pure behavior (excepting proportional factors).
Clearly the scaling has been deactivated, as expected, and $z\simeq 1$ as
corresponds to the behavior of the pure system. This phenomenon could never
happen with a system away from the quantum Griffiths zone.

In {\bf conclusion}, Quantum Monte Carlo calculations in diluted Ising
models in a transverse field at $T=0$ show that dynamic scaling holds above
the percolation threshold $(p>p_{c})$, at the percolation threshold $%
(p=p_{c})$ and inside the quantum Griffiths-McCoy phase $(p<p_{c})$. The
evolution of the activated scaling has been characterized, showing how it
grows monotonously towards the value corresponding to the percolation
threshold and how it appears to remain nearly constant in the Griffiths
zone for values of $p$ near $p_{c}$. A new way to deactivate the scaling in the Griffiths zone by thermal
dilution has been proposed. The expected deactivation works due to the fact
that phase transitions come from strongly coupled regions, and not from the
whole sample. The deactivation has been clearly confirmed by means of
Quantum Monte Carlo simulations.

We thank P. A. Serena for generous access to his computing facilities.
Financial support from DGCyT through grant PB96-0037 is gratefully
acknowledged.

\begin{figure}
\epsfysize=8cm\epsfxsize=\columnwidth\epsfbox{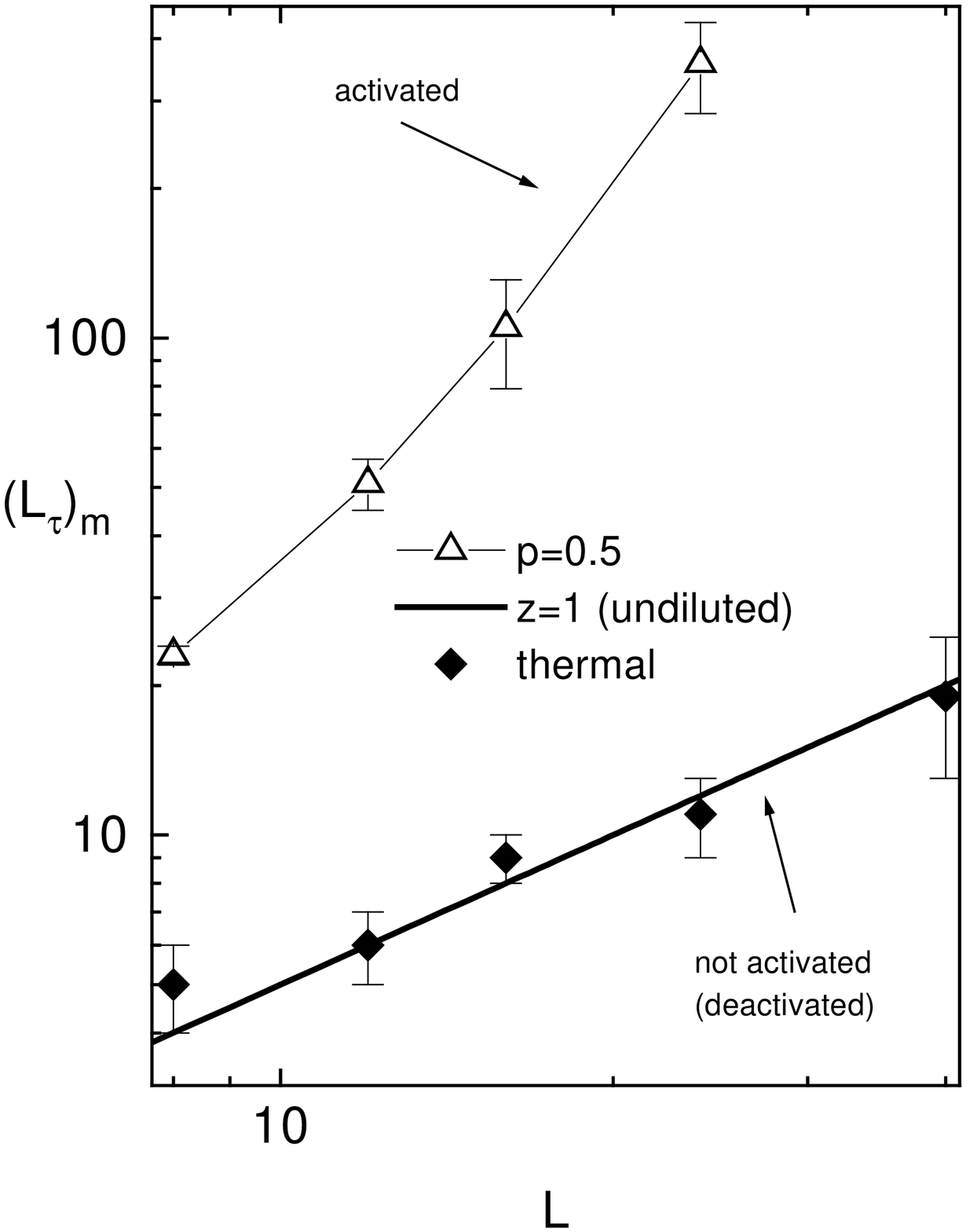}
\caption{$(L_{\tau })_{m}$ vs. $L$ for a randomly diluted system with spin concentration $p=0.5$
and a thermally diluted system.
The thin line is a guide for the eye, and the thick line represents the behavior
expected for the pure case ($z=1$) excepting proportional factors. Note how the scaling has been
deactivated for the thermal case.}
\label{fig4}
\end{figure}





\end{multicols}

\end{document}